\documentclass[twocolumn,showpacs,preprintnumbers,amsmath,amssymb]{revtex4}
\usepackage{amsfonts}
\usepackage{amsmath}
\usepackage{graphicx}
\usepackage{dcolumn}
\usepackage{bm}
\usepackage{overpic}
\usepackage{booktabs}

\begin{document}
\title{The reinforcing influence of recommendations on global diversification}
\author{An Zeng$^{1}$, Chi Ho Yeung$^{1}$, Mingsheng Shang$^{2}$, Yi-Cheng Zhang$^{1,2}$\footnote{yi-cheng.zhang@unifr.ch}}
 \affiliation{$^{1}$Department of Physics, University of Fribourg, Chemin du Mus\'{e}e 3, CH-1700 Fribourg, Switzerland\\
 $^{2}$Web Sciences Center, School of Computer Science and Engineering, University of Electronic Science and Technology of China Chengdu 610054, P. R. China
}

\date{\today}

\begin{abstract}
Recommender systems are promising ways to filter the overabundant information in modern society.
Their algorithms help individuals to
explore decent items,
but it is unclear how
they allocate popularity among items.
In this paper,
we simulate successive recommendations
and measure their influence on the dispersion of item popularity by Gini coefficient.
Our result indicates that
local diffusion and collaborative filtering
reinforce the popularity of
hot items, widening the popularity dispersion.
On the other hand,
the heat conduction algorithm
increases the popularity of the niche items
and generates
smaller dispersion of item popularity.
Simulations are compared to mean-field
predictions.
Our results suggest that recommender systems have reinforcing influence on global diversification.

\end{abstract}



\maketitle

\section{Introduction.}
Due to the rapid expanding of the internet,
we are overloaded by the unlimited information on the World Wide Web~\cite{Broder2000}.
For instance,
one has to choose among millions of candidate commodities to shop online.
Comprehensive exploration is infeasible~\cite{Maslov01}.
As a result,
various recommendation approaches have been proposed to help filtering
the relevant information~\cite{Adomavicius05,Cacheda11}.
For instance,
the
popularity-based recommendations (PR),
which recommend the most popular items to users, are commonly adopted in online recommender systems.
However, such recommendations
are not personalized
such that identical items are recommended for individuals with far
different taste.
By comparison, the collaborative filtering (CF)
makes use of collective data of individual preference
and
provides personalized recommendations~\cite{Konstan1997,Herlocker04}. So far, CF has been successfully applied to many online
applications.

Recently, recommendation algorithms have been proposed from a
physics perspective~\cite{Zhou07,Zhang07}.
For instance,
diffusion is applied on the user-item bipartite networks to explore items of potential interest for a user.
This mass diffusion (MD) algorithm is shown to outperform CF in the recommendation accuracy~\cite{Zhou07}.
However,
a similar problem as observed in PR is found in MD:
diffusion-based recommendations are biased to popular items even individual preferences are considered.
In fact, a good recommendation algorithm
should recommend items of personal interest and at the same time maximize the diversity of choices.

An alternative approach
based on the heat conduction (HC) on the user-item graphs
is thus introduced~\cite{Zhang07}. This method provides users with many novel items and leads to
diverse recommendation results among users.
However,
HC has low accuracy compared with MD.
The paradox is eventually solved by combining MD with HC in a hybrid algorithm~\cite{Zhou10},
which can be well-tuned to obtain significant improvement in both recommendation accuracy and
item diversity.

Though they are helpful in filtering information,
recommendation algorithms may impose reinforcing influence on the system, by guidance to one's choices which influence subsequent recommendations and hence choices of others. The influence is amplified with successive recommendations. We note that such perspective is employed to explain the evolution movie popularity~\cite{Chaos043101,Yeung11}, which yields consistent predictions compared with observed data. It is thus interesting to examine such influence on recommender systems. Unlike most existing works which are devoted to improving recommendation accuracy~\cite{Herlocker04}, our present study presents a physics perspective and utilizes microscopic interactions to explain and predict macroscopic behaviors of recommender systems~\cite{Shang09,Lambiotte2005}.

In this paper, we use the Gini coefficient to measure the dispersion in item popularity~\cite{Gini}.
We note that a small dispersion implies similar popularity among items,
and hence diverse recommendations for users.
We consider various conventional algorithms including
the popularity-based, the collaborative filtering, the mass diffusion and the heat conduction algorithms. We focus on the physical aspects and study numerically and theoretically the reinforcing influence of recommendations on the dispersion of item popularity.
The result indicates that MD and CF
reinforce the popularity of popular items,
as similar to PR.
On the other hand,
the heat conduction algorithm increases the popularity of the niche items
and generates smaller dispersion in item popularity.
Our results suggest that recommender systems have reinforcing influence on global diversification.

\section{Dispersion of Item Popularity}
We quantify the influence of recommender systems by measuring the changes in the dispersion of item popularity after successive recommendations. If the dispersion is large, some items dominate in popularity and users have limited choices. On the other hand, if the dispersion is small, items have similar popularity and users enjoy diverse recommendations.

To quantify such global diversity, we make use of Gini coefficient $G$~\cite{Gini} to measure the dispersion of item popularity, as in the case of individual wealth.
In addition to wealth, it has been used to measure dispersion in
sociology, science and engineering.
Mathematically,
it is given by
\begin{equation}
G=1-2\int_{0}^{1}C\left(x\right)dx,
\end{equation}
where $C(x)$ is the normalized cumulative popularity when items are ranked in ascending order of popularity,
with $x$ being the normalized rank. Specifically,
$G=0$ corresponds to uniform popularity among items,
while $G=1$ corresponds to maximal dispersion.

To see how Gini coefficient quantifies the changes in popularity dispersion
we study as examples the scientific citation data and the
baby name data. The scientific citation data is based on the citation relation in the APS (American Physics Society) journals from 1893 to 2009~\cite{APSdata}, and the
baby name data is based on the first names taken from US Social Security Administration, and contain the top 1000
boy and girl names every year from 1880 to 2009~\cite{babynames}. What we are interested most is how the dispersion changes with time in these two systems. The results are reported in
Fig. 1 from which we can see the Gini coefficient keeps increasing in
APS citation system while decreasing in baby name system.
Due to the technological advances,
ones get access to far more information than before.
Good papers can thus
have wider spread
and are cited more which leads to
a larger dispersion and hence lower global diversity.
Similarly, parents know more candidate names for babies,
and the system shows increasing diversification.

\begin{figure}
  \center
  \includegraphics[width=9cm]{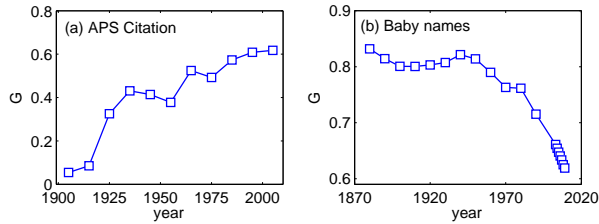}
\caption{(Color online) The change of Gini index with time in the APS citation system and the Baby name system.}\label{fig1}
\end{figure}

The above examples show that the changes in global diversity are well captured by the Gini coefficient. We thus make use of Gini coefficient to examine the influence of recommender systems on popularity dispersion.

\section{The reinforcing influence of recommender systems}

\begin{figure}
  \center
  \includegraphics[width=5cm]{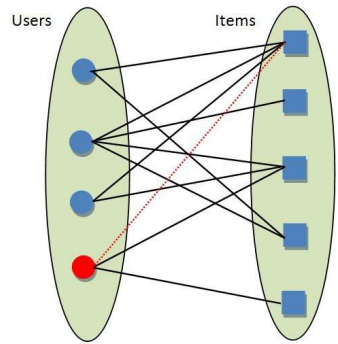}
\caption{(Color online) An illustrative example of the evolution of the bipartite network. The red node corresponds to the active user, and the red link corresponds to the choice made by the user according to recommendation results.
}\label{fig2}
\end{figure}

We investigate in this section the influence of recommender systems on the global diversity by examining dispersion in item popularity.
Here, we consider four recommendation algorithms including mass diffusion (MD), heat conduction (HC), user-based collaborative filtering (UCF), item-based collaborative filtering (ICF).
In addition, we consider two benchmark algorithms including popularity-based recommendation (PR) and random recommendation (RR),
corresponding to the recommendations of respectively most popular and random items.

We first give brief descriptions of the MD algorithm.
Consider a system of $N$ users and $M$ items represented by a bipartite network
with adjacency matrix $A$,
where the element $a_{i\alpha}=1$ if user $i$ has collected object $\alpha$, and $a_{i\alpha}=0$ otherwise (throughout this paper we
use Greek and Latin letters, respectively, for object- and user-related indices).

For a target user $i$,
the algorithm starts by
assigning one unit of resources to objects collected by $i$, and redistributes the resource through the user-item network.
We denote the vector $\textbf{f}$ as the initial resources on items where $f_{\alpha}$ is the resource possessed by object $\alpha$. The redistribution is represented by $\widetilde{\textbf{f}}=W\textbf{f}$, where
\begin{equation}
w_{\alpha\beta}=\frac{1}{k_{\beta}}\sum\limits_{l=1}^{M}\frac{a_{l\alpha}a_{l\beta}}{k_{l}},
\end{equation}
is the diffusion matrix, with $k_{\beta}=\sum^{N}_{i=1}a_{i\beta}$ and $k_{l}=\sum^{M}_{\gamma}a_{l\gamma}$ denoting the degree of object $\beta$ and user $l$ respectively. Technically, recommendations for a given user $i$ are obtained by setting
the initial resource vector $\textbf{f}^{i}$ in accordance with the objects the user has
already collected, that is, by setting $f^{i}_{\alpha}=a_{i\alpha}$. The resulting recommendation
list of uncollected objects is then sorted according to $\widetilde{f}^{i}_{\alpha}$ in descending order.
Physically, the diffusion is equivalent to a three-step random walk starting with $k_i$ units of resources on the target user $i$. The \emph{recommendation score} of an item is taken to be the resources on the item after the diffusion.
The scores for objects that user $i$ have already collected
are set to $0$. The recommendation list for user $i$ is generated by ranking all his/her uncollected objects in descending order of their final resources.

The HC algorithm works similar to the MD algorithm, but instead of a diffusion process, the scores are evaluated by a conduction process as represented by
\begin{equation}
w_{\alpha\beta}=\frac{1}{k_{\alpha}}\sum\limits_{l=1}^{M}\frac{a_{l\alpha}a_{l\beta}}{k_{l}}.
\end{equation}
Physically, the temperature of an object is considered to be the average temperature of its nearest neighborhood, i.e. its connected objects. The higher the temperature of an item is, the higher its recommendation score.

The CF algorithms provide recommendations based on user or item similarities.
It is divided into two main categories: the user-based CF and the item-based CF.
In UCF,
the recommendation score of an item is evaluated based on the similarity between the target user and the users who collected the item. The final recommendation score for each item can be written as
\begin{equation}
\widetilde{f}^{i}_{\alpha}=\sum^{N}_{j=1}s_{ij}a_{j\alpha}.
\end{equation}
where $s_{ij}$ is the similarity between user $i$ and $j$.

In ICF, the recommendation score of an item is evaluated based on its similarity with the collected items of the target user. Similarly, the final recommendation score for each item can be written as
\begin{equation}
\widetilde{f}^{i}_{\alpha}=\sum^{M}_{\beta=1}s_{\alpha\beta}a_{i\beta}.
\end{equation}
where $s_{\alpha\beta}$ is the similarity between item $\alpha$ and $\beta$.

The measure of similarities used in CF is subject to definition.
Here we define the measure of similarity as the number of common neighbors~\cite{CN1971} in the bipartite networks.

With the above mentioned algorithms, we consider a scenario of recommender systems as follows.
At every step a random user is selected as the active user, based on whom the recommendation scores of all items are evaluated. For simplicity, we assume that the active user would accept the recommendation results and select the uncollected item with the highest recommendation score, i.e. adding a link between the active user and the item in the bipartite network. An illustrative example is shown in Fig. 2. The red node corresponds to the active user, and the red link corresponds to the choice made by the user according to recommendation results.

In one \emph{marco-step} of our simulation, we randomly choose 10 percent of users as active users. After each macro-step, we evaluate the dispersion the item popularity by Gini coefficient. Note that we do not consider the growth of the system since introducing new users or items may involve the cold start problem for them~\cite{zico2010}. The datasets we will examine are the subsets of data obtained from four online systems: Movielens, Netflix, delicious and Amazon. These data are random samplings of the whole records of user activities in these websites, the descriptions of data are given in Table I.

\begin{table}[!htb]
 \extrarowheight=0.4em
 \tabcolsep=4pt
\begin{center}
\caption{Description of the data}
\begin{tabular}{lccclccc}

\hline
  network  &Users &Items  &Links &Sparsity\\

\hline
  Movielens &$943$ &$1,682$ &$82,520$ &$5.20\cdot10^{-2}$\\
  Netflix  &$3,000$ &$3,000$ &$197,248$ &$2.19\cdot10^{-2}$\\
  Delicious  &$1,000$ &$18,700$ &$63,290$ &$3.40\cdot10^{-3}$\\
  Amazon &$5,000$ &$12,377$ &$36,391$ &$5.88\cdot10^{-4}$\\
 \hline
\end{tabular}
\end{center}
\end{table}

We show in Fig. 3 the evolution of Gini coefficient in simulations as a function of macro-step. As we can see, the Gini coefficient increases in the presence of MD, UCF and ICF algorithms. This corresponds to their reinforcing influences on the system, leading to a wider dispersion of item popularity after successive recommendations. A further evidence can be seen in Fig. 4, which shows that popular items become more popular, while neglecting the rest of the items. This corresponds to an undesired influence, as choices and visions for users become more limited in the presence of these recommendation algorithms.

We can further understand the reinforcing influences of the MD, UCF and ICF recommender systems by comparing their Gini coefficients with the unpersonalized popularity-based algorithm. As shown in Fig. 3, similar trends are observed between the four algorithms. We can examine the underlying reasons in Fig. 4, which shows that the MD, UCF, ICF and PR algorithms only recommend to users the most popular items. These results imply that the changes in the distribution of item popularity are similar in these four algorithms. Therefore, personalized elements in the MD, UCF and ICF algorithms do not increases the global diversity as compared to the unpersonalized PR algorithm.

\begin{figure}
  \center
  \includegraphics[width=9cm]{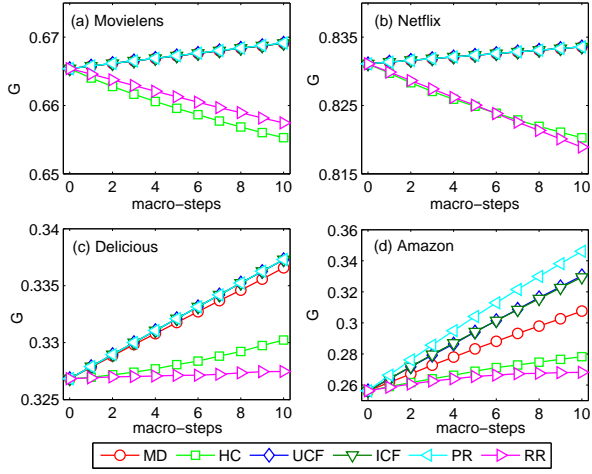}
\caption{(Color online) The change of the Gini coefficient for items' popularity when using different recommendation methods in real systems. The results are averaged on 100 independent realizations.}\label{fig3}
\end{figure}

\begin{figure}
  \center
  \includegraphics[width=9cm]{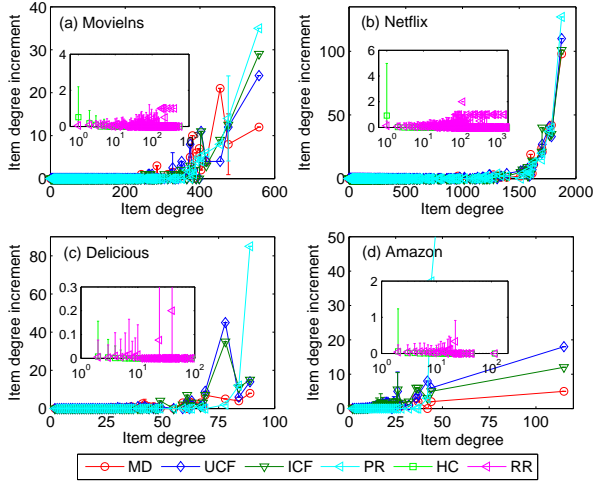}
\caption{(Color online) The items popularity increment when using different recommendation method in real systems. The results are averaged on 100 independent realizations.}\label{fig4}
\end{figure}

On the other hand, the HC algorithm behaves quite differently from the other algorithms. As we can see in Fig. 3, it generally decreases the Gini coefficient in Movielens and Netflix, where density of links is high. In sparse systems, the three-step conduction process can only reach some items with large degree, and inevitably add links to hot items. This leads to an increasing Gini coefficient. Moreover, the HC method is also different from the random recommendations as we can see from Fig. 4. Instead of uniform addition of links, it inclined to add links to items with small degree. It implies that the HC algorithm does not reinforce the popularity of hot items as the MD and CF algorithms.

\section{The Mean-field Approximation}

To better understand their influences of recommender systems, we derive analytically the distribution of item scores after the recommendation processes. The major difficulty in analysis comes from the particular network topology of each dataset, which embeds the non-trivial correlations between users and items~\cite{NewmanPRL,JSM2006}. Here we focus on the recommendation influences, and assume a simple topology where users and items are randomly connected~\cite{NewmanPRE}. This corresponds to a crude mean-field approximation, but such assumption facilitates the analysis and the illustration of physical behaviors underlying the recommendation algorithms.

To begin our analysis, we derive the probability $p_{i\alpha}$ that a user $i$ and an item $\alpha$ are connected in a random graph. Suppose we start with $k_{i}$ cavities on user $i$ and $k_{\alpha}$ on item $\alpha$, which are respectively the degree of $i$ and $\alpha$.
If one cavity is picked randomly among the items, the probability that $\alpha$ being picked is $\frac{k_{\alpha}}{\sum^{M}_{\beta=1}k_{\beta}}$. It implies that $p_{i\alpha}=1-(1-\frac{k_{\alpha}}{\sum^{M}_{\beta=1}k_{\beta}})^{k_{i}}$, where $(1-\frac{k_{\alpha}}{\sum^{M}_{\beta=1}k_{\beta}})^{k_{i}}$ is the probability that $i$ is not connected to $\alpha$. As $\sum^{M}_{\beta=1}k_{\beta} \gg k_{\alpha}$, expansion to the first order of  $k_{\alpha}$ leads to $p_{i\alpha}\approx 1-(1-k_{i}\frac{k_{\alpha}}{\sum^{M}_{\beta=1}k_{\beta}})=\frac{k_{i}k_{\alpha}}{c}$, where $c=\sum^{M}_{\beta=1}k_{\beta}$ is the total number of links in the bipartite network.

We then derive the mean-field expression of recommendation scores in the MD recommender system. As mentioned above, the MD method is based on the three-step diffusion. The resource vector for items in the first step and last step are denoted respectively by $\textbf{f}$ and $\widetilde{\textbf{f}}$. In the second step, the resources are in users' side and the corresponding vector is denoted as $\textbf{e}$. By considering the last step of the diffusion process, the score of $\alpha$ from user $i$ is given by $\widetilde{f}^{i}_{\alpha}=(1-p_{i\alpha})\sum^{N}_{j=1}\frac{e^{i}_{j}p_{j\alpha}}{k_{j}}$.
Substitution of $p_{j\alpha}=\frac{k_{j}k_{\alpha}}{c}$ leads to
\begin{equation}
\widetilde{f}^{i}_{\alpha}=\left(1-\frac{k_{i}k_{\alpha}}{c}\right)\sum^{N}_{j=1}\left(\frac{e^{i}_{j}}{k_{j}}\frac{k_{j}k_{\alpha}}{c}\right)=\left(1-\frac{k_{i}k_{\alpha}}{c}\right)\frac{k_{i}k_{\alpha}}{c},
\end{equation}
as $\sum e^{i}_{j}=\sum f^{i}_{\alpha}=k_{i}$.

Next we derive the scores for the HC algorithms by again considering the last step of the conduction process. However, the total ``resources" does not conserve in heat conduction but instead the temperature of user $j$ is given by $e^{i}_{j}=\frac{k_{i}}{M}\sum_{\gamma=1}^{M}\frac{p_{j\gamma}}{k_{j}}=\frac{k_{i}}{M}$, where $\frac{k_{i}}{M}$ corresponds to the random choices of initial collected item for $i$. Therefore,
\begin{equation}
\widetilde{f}^{i}_{\alpha}=\left(1-\frac{k_{i}k_{\alpha}}{c}\right)\sum^{N}_{j=1}\left(\frac{k_{i}}{Mk_{\alpha}}\frac{k_{j}k_{\alpha}}{c}\right)=\left(1-\frac{k_{i}k_{\alpha}}{c}\right)\frac{k_{i}}{M}
\end{equation}

In the user-based CF, scores of items are evaluated by the similarity between the target user and the users who have collected it.
The user similarity is given by the number of common neighbors. Therefore,
$\widetilde{f}^{i}_{\alpha}=(1-p_{i\alpha})\sum^{N}_{j}s_{ij}p_{j\alpha}$ where
$s_{ij}\approx\frac{k_{i}k_{j}}{M}$ in the mean-field approximation. The score for object $\alpha$ is then approximated by
\begin{equation}
\widetilde{f}^{i}_{\alpha}=\left(1-\frac{k_{i}k_{\alpha}}{c}\right)\sum^{N}_{j=1}\left(\frac{k_{i}k_{j}}{M}\frac{k_{j}k_{\alpha}}{c}\right)=\left(1-\frac{k_{i}k_{\alpha}}{c}\right)\frac{k_{i}k_{\alpha}b}{cM}.
\end{equation}
where $b=\sum^{N}_{j=1}k_{j}^{2}$ is a constant for a given network.

As similar to user-based CF, the item similarity in item-based CF can be approximated by $s_{\alpha\beta}=\frac{k_{\alpha}k_{\beta}}{N}$ in the mean-field approximation. The score for object $\alpha$ is then approximated by
\begin{equation}
\widetilde{f}^{i}_{\alpha}=\left(1-\frac{k_{i}k_{\alpha}}{c}\right)\sum^{M}_{\beta=1}\left(\frac{k_{i}k_{\beta}}{c}\frac{k_{\alpha}k_{\beta}}{N}\right)=\left(1-\frac{k_{i}k_{\alpha}}{c}\right)\frac{k_{i}k_{\alpha}d}{cN}.
\end{equation}
where $d=\sum^{M}_{\beta=1}k_{\beta}^{2}$ is a constant for a given network.

In order to compare the simulated results and the mean-field predictions, we evaluate the corresponding total scores $F_{\alpha}=\sum_{i}^{N}\widetilde{f}^{i}_{\alpha}$ that a item receives from all the users. As shown in Fig. 5, the mean-field approximation captures both the magnitude and the trend of the recommendation scores.

Further insights are drawn by noting $c \gg k_{i}k_{\alpha}$ in most systems, which implies $\widetilde{f}^{i}_{\alpha} \propto k_{i}k_{\alpha}$ in Eqs. (4), (5) and (7).
Since we assume that users always accept the item with highest recommendation scores, the recommendation scores in the MD, UCF and ICF cases are thus similar to the PR algorithm which recommends the most popular items.
This again shows the reinforcing influence of these recommendation algorithms. On the other hand, Eq. (5) suggests $\widetilde{f}^{i}_{\alpha} \propto k_{i}$ in the HC algorithm,
which is item independent as in the case of random recommendations.

Though the approximated scores of HC agree well with RR, their behaviors are difference in terms of choices of items. According to $\widetilde{f}^{i}_{\alpha}=(1-p_{i\alpha})\sum^{N}_{j=1}\frac{k_{i}p_{j\alpha}}{k_{j}k_{\alpha}}$, users select the reachable items with lowest degree after three-step conduction, compared to the random choice. Therefore, the HC and RR algorithms show different influence on the dispersion of item popularity, as we can see in Fig. 3(c) and (d).

\begin{figure}
  \center
  \includegraphics[width=9cm]{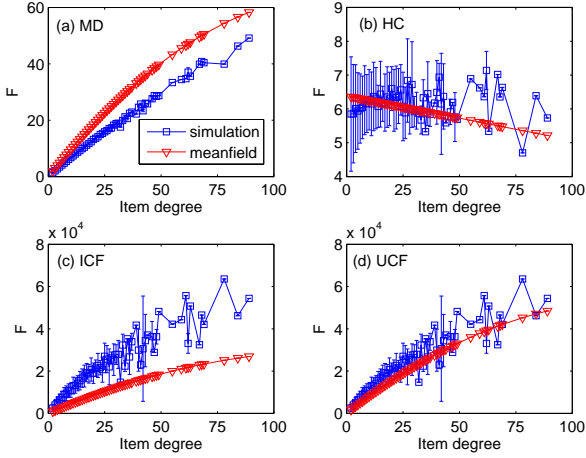}
\caption{(Color online) The simulation result and the theoretical result of the total recommendation score versus the original item degree in different recommendation engines. The simulation results are averaged on 100 independent realizations.}\label{fig5}
\end{figure}

\section{Steady Gini Coefficient by Hybrid Recommendations}

As we have seen from the previous sections, the MD algorithm reinforces the popularity of hot items and limits available choices, while the HC algorithm recommends items with low popularity and increases global diversity. It is thus interesting to examine the influence on diversity if these two algorithms with opposite influences are combined. We thus adopt the hybrid algorithm of MD and HC proposed in~\cite{Zhou10}, with the new recommendation score $\widetilde{h}_{\alpha}$ given by
\begin{equation}
\label{eq_hybrid}
\widetilde{h}_{\alpha}=\lambda\widetilde{f}^{\rm MD}_{\alpha}+\left(1-\lambda\right)\widetilde{f}^{\rm HC}_{\alpha}.
\end{equation}
The parameter $\lambda$ adjust the relative weight between the two algorithms. When $\lambda$ increases from $0$ to $1$, the hybrid algorithm change gradually from HC to MD. We remark that though Eq. (\ref{eq_hybrid}) corresponds to a linear combination of scores, the hybrid algorithm is a non-linear combination of HC and MD as users select only items with highest scores.

\begin{figure}
  \center
  \includegraphics[width=9cm]{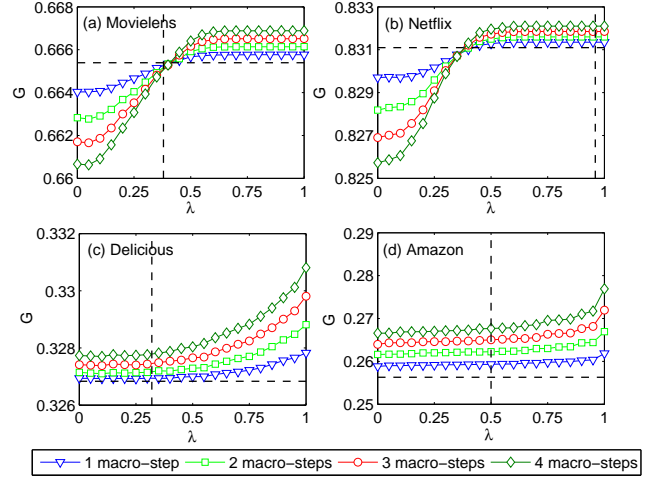}
\caption{(Color online) The change of the Gini coefficient for items' popularity when adjusting the $\lambda$ in the hybrid recommendation algorithm in real systems. The results are averaged on 100 independent realizations.}\label{fig6}
\end{figure}

The influence of the hybrid algorithm on Gini coefficient is shown in Fig. 6 as a function of $\lambda$. The lines with different symbols correspond to Gini coefficient measured after increasing macro-step. As we can see from Fig. 6 (a) and (b), the Gini coefficient increases with $\lambda$, corresponds to a transition from HC to MD recommender systems. It is interesting to note that Gini coefficient shows a significant increase in a short range of $\lambda$ on the Netflix and Movielens datasets, and becomes saturated afterwards. The saturated Gini coefficient corresponds to dominance of the MD algorithm such that only popular items are recommended, despite the presence of HC algorithm. Similar behaviors are not observed in Fig. 6 (c) and (d) in the Delicious and Amazon datasets, which are sparse compared to the Netflix and Movielens datasets.

Another interesting behavior is noted in Fig. 6 (a) and (b) when we compare the Gini coefficient after different macro-steps of recommendations. As we can see, the lines with different symbols intersect at a particular value of $\lambda$, suggesting a steady Gini coefficient after the reinforcement of recommendations. The corresponding value of $\lambda$ thus corresponds to the balance between the HC and MD algorithms, leading to steady dispersion in item popularity. This is desirable when one considers the reinforcing influence on global diversity as undesired side-effect of recommender systems. These values of $\lambda$ and the corresponding Gini coefficient are compared respectively to the values of $\lambda$ with optimal recommendation accuracy \cite{Herlocker04} and the Gini coefficient before recommendation algorithms are implemented. These results show that high recommendation accuracy does not always guarantee a global diversity, leading to a paradox in recommendations.

\section{Conclusion}
Recommendation is an effective way to solve the problem of excess information. However, it is unclear how
they allocate popularity among items. In this paper,
we simulate successive recommendations
and measure their influence on the dispersion of item popularity by Gini coefficient.
Our result indicates that
local diffusion and collaborative filtering
reinforce the popularity of
hot items, widening the popularity dispersion.
On the other hand,
the heat conduction algorithm
increases the popularity of the niche items
and generates
smaller dispersion of item popularity.
Simulations are compared to mean-field
approximation.
Our results indicate that there is reinforcing influence of recommender systems on global diversification.
This work provides a deeper understanding of these recommendation methods, highlights the importance of the global diversity and may shine some light for developing a new recommendation method that can directly controls the global diversity.

\section*{Acknowledgement.}
This work was partially supported by the Swiss National Science Foundation under Grant No. (200020-132253), the National Natural Science Foundation of China(Grant No. 60973069) and the Sichuan Provincial Science and Technology Department(Grant No.2010HH0002).

\end{document}